\documentclass[a4paper, preprint, preprintnumbers, superscriptaddress, showpacs, showkeys, nofootinbib, floatfix, twocolumn, 10pt]{revtex4-1}
\usepackage[utf8]{inputenc}
\usepackage[english]{babel}
\usepackage{url}
\usepackage{amsthm,amssymb,amsmath,mathtools}
\usepackage{physics,bbm,enumitem}

\allowdisplaybreaks

\begin{document}

\title[LAQC BD states comments]{Comment on `Local available quantum correlations for Bell Diagonal states and markovian decoherence'}

\author{David M. Bellorin R.}
\affiliation{Departamento de Física, Universidad Simón Bolívar, AP 89000, Caracas 1080, Venezuela.}

\author{Hermann L. Albrecht Q.}
\email[Corresponding Author: ]{albrecht@usb.ve}
\affiliation{Departamento de Física, Universidad Simón Bolívar, AP 89000, Caracas 1080, Venezuela.}

\begin{abstract}
In this brief paper, we complete the analysis presented previously in [RMF 64 (2018) 662–670] regarding the quantifiers of the classical correlations and the so-called local available quantum correlations for Bell Diagonal states. A correction is introduced in their previous expressions once two cases within the optimizations are included.
\end{abstract}
\preprint{SB/F/493-21}

\maketitle

\section{Introduction}
In \cite{LAQC_BD}, the analytical results of the correlation quantifiers related to the so-called local available quantum correlations (LAQC) \cite{LAQC} for the family of Bell Diagonal states \cite{Horodecki-BD_states} were presented. These states are written in the Bloch representation as
\begin{equation}\label{eq:rho_BD}
\rho^{BD}=\frac{1}{4}\left(\mathbbm{1}_4 +\sum_{i=1}^{3} c_{i}\sigma_i\otimes\sigma_i\right),
\end{equation}
where the coefficients $c_i\in[-1,1]$ are such that $\rho^{BD}$ is a well-behaved density matrix (i. e. has non-negative eigenvalues) and $\sigma_i$ are the well-known Pauli matrices, i.e. $\sigma_1=\smqty(\pmat{1})$, $ \sigma_2=\smqty(\pmat{2})$, and  $\sigma_3=\smqty(\pmat{3})$. As is readily seen from \eqref{eq:rho_BD}, BD states have null local Bloch vectors, and their reduced matrices $\rho_{A(B)}$ are maximally mixed. That is, they are proportional to identity,
\begin{equation}
    \rho_A=\rho_B = \frac{1}{2}\,\mathbbm{1}_2.
\end{equation}

In their 2015 paper, Mundarain and Ladrón de Guevara introduced a new type of quantum correlations defined in the complementary basis of an optimal computational one. To do so, they establish a generic orthonormal basis
\begin{eqnarray}\label{eq:BaseOrtonormalGen}
\ket{\mu_0^{(n)}} &=& \cos\qty(\frac{\theta_n}{2})\ket{0^{(n)}} +\sin\qty(\frac{\theta_n}{2})\exp^{i\phi_n}\ket{1^{(n)}},\\
\ket{\mu_1^{(n)}} &=& -\sin\qty(\frac{\theta_n}{2})\ket{0^{(n)}} +\cos\qty(\frac{\theta_n}{2})\exp^{i\phi_n}\ket{1^{(n)}},\nonumber
\end{eqnarray}
\noindent{}where $n=1$ denotes subsystem A and $n=2$ subsystem B, respectively. The optimal basis is fixed by minimizing the conditional entropy \begin{equation}\label{eq:S(rho||X)}
S\qty(\rho_{AB}||X_\rho)=\min_{\Omega_c}S\qty(\rho_{AB}||\chi_\rho)\,,
\end{equation}
\noindent{}where $S\qty(\rho_{AB}||\chi)=-\mathrm{Tr}(\rho\mathrm{log}_2\chi)-S(\rho_{AB})$ is the relative entropy and
\begin{equation}\label{eq:Chi_rho}
\begin{aligned}
\chi_\rho =&\sum_{ij}R_{ij}\,\dyad{\mu_i^{(1)},\mu_j^{(2)}}\,,\\ 
R_{ij} =& \expval{\rho_{AB}}{\mu_i^{(1)},\mu_j^{(2)}}\,.
\end{aligned}
\end{equation}
The classical correlations quantifier defined in this context in \cite{LAQC} is given by 
\begin{equation}\label{eq:CorrClasicas}
\mathcal{C}(\rho_{AB}) \equiv S\left(X_{\rho_{AB}}||\Pi_{_{X_{\rho_{AB}}}}\right),
\end{equation}
\noindent{}where  $\Pi_{_{X_{\rho_{AB}}}}$ is the product state nearest to $X_{\rho_{AB}}$.

As was shown by Modi et al. \cite{Modi-RelativeEntropy}, the relative entropy in the above expression can be written in terms of the Mutual Information, given by \begin{equation}\label{eq:InfoMutua}
I(\rho_{AB})\equiv S\qty(\rho_A)+S\qty(\rho_B)-S\qty(\rho_{AB}),
\end{equation}
of state $X_{\rho_{AB}}$. Moreover, since the mutual information \eqref{eq:InfoMutua} may be written as
\begin{equation}\label{eq:Info_Mutua-Prob}
I\qty(\rho_{AB}) = \sum_{i,j} P(i_A,j_B)\log_2\qty[\frac{P(i_A,j_B)}{P(i_A)P(j_B)}],
\end{equation}
where $P(i_A,j_B) = \expval{\rho_{AB}}{\mu_i^{(1)},\mu_j^{(2)}}$ are the probability distributions corresponding to $\rho_{AB}$, and $P(i_A) = \expval{\rho_A}{\mu_i^{(1)}}$, $P(j_B) = \expval{\rho_B}{\mu_i^{(2)}}$ the ones corresponding to its marginals $\rho_A$ and $\rho_B$, the classical correlations quantifier \eqref{eq:CorrClasicas} can be written in terms of the $R_{ij}\qty(\theta_A,\phi_A,\theta_B,\phi_B)$ coefficients as
\begin{equation}\label{eq:CorrClas-Rij}
\begin{aligned}
    \mathcal{C}\qty(\rho_{AB}) &= \min_{\smqty{\theta_A, \phi_A\\ \theta_B, \phi_B}} \Bigg\{\sum_{i,j} R_{ij}\qty(\theta_A,\phi_A,\theta_B,\phi_B)\\
    &\qq{ } \times\log_2\qty[\frac{R_{ij}\qty(\theta_A,\phi_A,\theta_B,\phi_B)}{R_{i}\qty(\theta_A,\phi_A)R_{j}\qty(\theta_B,\phi_B)}]\Bigg\}.
\end{aligned}
\end{equation}

The above minimization over the angles $\theta_A$, $\phi_A$, $\theta_B$, and $\phi_B$ defines a new computational basis $\qty{\ket{0^{(m)}}_{opt},\ket{1^{(m)}}_{opt}}$ relative to $\qty{\ket{\mu_i^{(n)}}}$ \eqref{eq:BaseOrtonormalGen}. The state $\rho_{AB}$ is rewritten, and the complementary one is then defined in terms of a new general orthonormal basis:
\begin{eqnarray}\label{eq:u0-u1}
  \ket{u_0^{(m)}}&=&\frac{1}{\sqrt{2}}\qty(\ket{0^{(m)}}_{opt} +\exp^{i\Phi_m}\ket{1^{(m)}}_{opt})\,,\nonumber\\
  \ket{u_1^{(m)}}&=&\frac{1}{\sqrt{2}}\qty(\ket{0^{(m)}}_{opt} -\exp^{i\Phi_m}\ket{1^{(m)}}_{opt})\,.
\end{eqnarray}
\noindent{}The corresponding probability distributions, \begin{equation}\label{eq:P_phi}
  P_\Phi(i_A,j_B,\Phi_1,\Phi_2) = \expval{\tilde{\rho}_{AB}}{u_i^{A},u_j^{B}},
\end{equation}
where $\tilde{\rho}_{AB}$ is the density matrix ${\rho}_{AB}$ written in the optimal computational basis, and the marginal probability distributions are determined. The maximization of $I\qty(\Phi_1,\Phi_2)$ \eqref{eq:Info_Mutua-Prob} corresponds to the LAQC quantifier:
\begin{equation}\label{eq:LAQC-quant}
    \mathcal{L}(\rho_{AB}) \equiv \max_{\qty{\Phi_1,\Phi_2}} I\qty(\Phi_1,\Phi_2)\,.
\end{equation}

\section{\label{sec:LAQCs-BD}Local available quantum correlations of Bell diagonal states}

Since Bell diagonal (BD) states have null local Bloch vector, it is straightforward that they are invariant under subsystem exchange $\vb{A}\leftrightarrow{}\vb{B}$. Therefore, the parametrization of the orthogonal basis  $\qty{\ket{\mu_i^{(n)}}}$ \eqref{eq:BaseOrtonormalGen} needs to respect this symmetry. As a consequence, only two angles, $\theta$ and $\phi$, are necessary to parametrize \eqref{eq:BaseOrtonormalGen}. With this taken into account, the coefficients $R_{ij}(\theta,\phi)$ are given by
\begin{eqnarray}\label{eq:Rij-BD_theta_1}
\begin{aligned}
R_{ij}(\theta,\phi)&= \frac{1}{4}\qty[1+(-1)^{i+j} c_3]\\
& +(-1)^{i+j}\frac{1}{2}\cos^2\qty(\frac{\theta}{2})\sin^2\qty(\frac{\theta}{2})\\
&\times\qty\big[(c_1+c_2)+\cos(2\phi)(c_1-c_2)-2c_3],
\end{aligned}
\end{eqnarray}
with $R_{00}(\theta,\phi)=R_{11}(\theta,\phi)$, $R_{01}(\theta,\phi)=R_{10}(\theta,\phi)$, and $R_i=\frac{1}{2}$. The minimization in \eqref{eq:CorrClas-Rij} leads to three different cases:
\begin{enumerate}[label=\Roman*.)]
\item For $\theta=0$ and $\phi=0$:
\begin{equation}
R_{00}(0,0) =\frac{1}{4}\qty(1+c_3)\qc R_{01}(0,0) =\frac{1}{4}\qty(1-c_3).
\end{equation}
\item For $\theta=\frac{\pi}{2}$ and $\phi=0$:
\begin{equation}
R_{00}\qty(\frac{\pi}{2},0) =\frac{1}{4}\qty(1+c_1)\qc
    R_{01}\qty(\frac{\pi}{2},0) =\frac{1}{4}\qty(1-c_1).
\end{equation}
\item For $\theta=\frac{\pi}{2}$ and $\phi=\frac{\pi}{2}$:
\begin{equation}
R_{00}\qty(\frac{\pi}{2},\;\frac{\pi}{2}) =\frac{1}{4}\qty(1+c_2),
    R_{01}\qty(\frac{\pi}{2},\frac{\pi}{2}) =\frac{1}{4}\qty(1-c_2).
\end{equation}
\end{enumerate}
Therefore, by defining 
\begin{equation}
    c_m\equiv \min\qty\big{\abs{c_1},\abs{c_2},\abs{c_3}},
\end{equation}
we can write the classical correlations quantifier \eqref{eq:CorrClas-Rij} as
\begin{equation}\label{eq:Corr_Clas-BD}
\begin{aligned}
\mathcal{C}\qty(\rho^{BD}) =& \frac{1+c_m}{2}\log_2(1+c_m)\\
 &+\frac{1-c_m}{2}\log_2(1-c_m).
\end{aligned}
\end{equation}

The above expression is the same as eq. (33) in \cite{LAQC_BD} but now the minimization achieved for $\theta=\frac{\pi}{2}$ and $\phi=0$ when $c_m=\abs{c_1}$ has been included.

To determine the LAQC quantifier \eqref{eq:LAQC-quant}, we have to rewrite the density matrix $\rho^{BD}$ in the optimal computational basis. Contrary to what is stated in \cite{LAQC_BD}, the density matrix of BD states does not remain invariant when written in the optimal computational basis. That is only true for Werner \cite{Werner}  and Werner-like states \cite{Werner-Like_Ghosh, Werner-Like_Verstraete}.

The density matrix $\tilde{\rho}^{BD}$ and their corresponding $P(i,j,\Phi)$ for each $\theta$ and $\phi$, with $P(0,0,\Phi)=P(1,1,\Phi)$, $P(0,1,\Phi)=P(1,0,\Phi)$, and $P(i,\Phi)=\frac{1}{2}$, are the following:
\begin{enumerate}[label=\Roman*.)]
\item For $\theta=0$ and $\phi=0$:
\begin{equation}
    \tilde{\rho}^{BD} =\frac{1}{4} \mqty(1+c_3 &0&0& c_1-c_2\\ 0&1-c_3&c_1+c_2&0\\ 0&c_1+c_2&1-c_3&0\\ c_1-c_2 &0&0& 1+c_3)
\end{equation}
and
\begin{equation}
\begin{aligned}
P(0,0,\Phi) &= \frac{1}{4}\qty[1+\frac{c_1+c_2}{2}+\frac{c_1-c_2}{2}\cos(2\Phi)],\\
P(1,0,\Phi) &= \frac{1}{4}\qty[1-\frac{c_1+c_2}{2}-\frac{c_1-c_2}{2}\cos(2\Phi)].
\end{aligned}
\end{equation}
\item For $\theta=\frac{\pi}{2}$ and $\phi=0$:
\begin{equation}
    \tilde{\rho}^{BD} =\frac{1}{4} \mqty(1+c_1 &0&0& c_3-c_2\\ 0&1-c_1&c_3+c_2&0\\ 0&c_3+c_2&1-c_1&0\\ c_3-c_2 &0&0& 1+c_1)
\end{equation}
and
\begin{equation}
\begin{aligned}
P(0,0,\Phi) &= \frac{1}{4}\qty[1+\frac{c_3+c_2}{2} +\frac{c_3-c_2}{2}\cos(2\Phi)],\\
P(1,0,\Phi) &= \frac{1}{4}\qty[1-\frac{c_3+c_2}{2} -\frac{c_3-c_2}{2}\cos(2\Phi)].
\end{aligned}
\end{equation}
\item For $\theta=\frac{\pi}{2}$ and $\phi=\frac{\pi}{2}$:
\begin{equation}
    \tilde{\rho}^{BD} =\frac{1}{4} \mqty(1+c_2 &0&0& c_3-c_1\\ 0&1-c_2&c_3+c_1&0\\ 0&c_3+c_1&1-c_2&0\\ c_3-c_1 &0&0& 1+c_2)
\end{equation}
and
\begin{equation}
\begin{aligned}
P(0,0,\Phi) &= \frac{1}{4}\qty[1+\frac{c_3+c_1}{2} +\frac{c_3-c_1}{2}\cos(2\Phi)],\\
P(1,0,\Phi) &= \frac{1}{4}\qty[1-\frac{c_3+c_1}{2} -\frac{c_3-c_1}{2}\cos(2\Phi)].
\end{aligned}
\end{equation}
\end{enumerate}

For each $\theta$ and $\phi$, $\Phi$ depends on $\abs{c_1}>\abs{c_2}$, $\abs{c_2}>\abs{c_3}$, or $\abs{c_1}>\abs{c_3}$, respectively. Therefore, as was done with the classical correlations quantifier \eqref{eq:Corr_Clas-BD}, defining  
\begin{equation}
    c_M\equiv \max\qty{\abs{c_1},\abs{c_2},\abs{c_3}}
\end{equation}
allows us to write a general expression for the LAQC quantifier that encompasses all these possibilities:
\begin{equation}\label{eq:LAQCs-BD}
\begin{aligned}
\mathcal{L}\qty(\rho^{BD}) =& \frac{1+c_M}{2}\log_2(1+c_M)\\
&\qq{} +\frac{1-c_M}{2}\log_2(1-c_M).
\end{aligned}
\end{equation}

As with the classical correlations quantifiers, the above expression is equivalent to the one presented in eq. (36) of \cite{LAQC_BD}. Nevertheless, this newly defined $c_M$ also includes $\abs{c_3}$. The case of $c_M=\abs{c_3}$ arises when the density matrix ${\rho}^{BD}$ is written in the optimal computational basis with $\theta = \frac{\pi}{2}$.

\section{Conclusions}
In this brief paper, we have completed the previous results regarding the so-called local available quantum correlations (LAQC) for Bell diagonal (BD) states. By including the cases of $\theta=\frac{\pi}{2}$ and $\phi=0$ in the minimization for determining the classical correlations, as well as the transformation of $\rho^{BD}$ when the optimal computational basis has $\theta=\frac{\pi}{2}$ we extended the definitions of $c_m$ and $c_M$ to include $\abs{c_1}$ and $\abs{c_3}$, respectively.

\section{Acknowledgments}
This work was partially funded by the \emph{2020 BrainGain Venezuela} grant awarded to H. Albrecht by the \emph{Physics without Frontiers} program of the ICTP. The authors would like to thank the support given by the research group GID-30, \emph{Teoría de Campos y Óptica Cuántica}, at the Universidad Simón Bolívar, Venezuela.  The authors would like to also thank D. Mundarain, from the Universidad Católica del Norte, Chile, as well as M.I. Caicedo and J. Stephany, from Universidad Simón Bolívar, Venezuela, for their comments and suggestions.

\bibliographystyle{unsrt}
\bibliography{biblio-qit}

\begin{thebibliography}{1}

\bibitem{LAQC_BD}
H.L. Albrecht~Q., M.I. Caicedo~S., and D.F. Mundarain.
\newblock Local available quantum correlations for bell diagonal states and
  markovian decoherence.
\newblock {\em Rev. Mex. Fís.}, 64:662–--670, Nov-Dec 2018.
\newblock arXiv:1803.02426.

\bibitem{LAQC}
D.F. Mundarain and M.L.L. de~Guevara.
\newblock Local available quantum correlations.
\newblock {\em Quantum Inf Process}, 14:4493–--4510, 10 2015.

\bibitem{Horodecki-BD_states}
Ryszard Horodecki and Michal/ Horodecki.
\newblock Information-theoretic aspects of inseparability of mixed states.
\newblock {\em Phys. Rev. A}, 54:1838--1843, Sep 1996.

\bibitem{Modi-RelativeEntropy}
K.~Modi, T.~Paterek, W.~Son, V.~Vedral, and M.~Williamson.
\newblock Unified view of quantum and classical correlations.
\newblock {\em Phys. Rev. Lett.}, 104:080501, Feb 2010.
\newblock arXiv:0911.5417.

\bibitem{Werner}
R.F. Werner.
\newblock Quantum states with einstein-podolsky-rosen correlations admitting a
  hidden-variable model.
\newblock {\em Phys. Rev. A}, 40:4277--4281, Oct 1989.

\bibitem{Werner-Like_Ghosh}
Sibasish Ghosh, Guruprasad Kar, Aditi Sen(De), and Ujjwal Sen.
\newblock Mixedness in the bell violation versus entanglement of formation.
\newblock {\em Phys. Rev. A}, 64:044301, Sep 2001.

\bibitem{Werner-Like_Verstraete}
Tzu-Chieh Wei, Kae Nemoto, Paul~M. Goldbart, Paul~G. Kwiat, William~J. Munro,
  and Frank Verstraete.
\newblock Maximal entanglement versus entropy for mixed quantum states.
\newblock {\em Phys. Rev. A}, 67:022110, Feb 2003.

\end{thebibliography}
\include{biblio-qit}

\end{document}